# Probing a Supersymmetric Model for Neutrino Masses at Ultrahigh Energy Neutrino Telescopes

M. Hirsch,[1] D. P. Roy,[1,2] and J. W. F. Valle[1]

[1]AHEP Group, Instituto de Fisica Corpuscular (IFIC), CSIC-U. de Valencia, Edificio de Instituto de Paterna, Apartado de Correos 22085, E-46071 Valencia, Spain

[2]Homi Bhabha Centre for Science Education, Tata Institute of Fundamental Research, Mumbai – 400088, India

**Abstract**

A bilinear R-Parity breaking SUSY model for neutrino mass and mixing predicts the lightest superparticle to decay mainly into a pair of tau leptons or b quarks along with a neutrino for relatively light SUSY spectra. This leads to a distinctive triple bang signature of SUSY events at ultrahigh energy neutrino telescopes like IceCube or Antares. While the expected signal size is only marginal at IceCube, it will be promising for a future multi-km$^3$ size neutrino telescope.

There is a good deal of current interest in the R-Parity breaking SUSY models of neutrino mass and mixing, since they simultaneously provide a solution to the hierarchy problem of the standard model [1]. Besides they are amenable to direct experimental test in the foreseeable future unlike the canonical see-saw models. In particular the bilinear R-Parity breaking SUSY model has relatively few RPB parameters, which can be fixed in terms of the observed neutrino masses and mixing angles [2-6]. Therefore it offers a predictive and well motivated extension of the minimal supersymmetric standard model (MSSM). It is described by the superpotential

$$W = W_{MSSM} + \varepsilon_i \hat{L}_i \hat{H}_u, \tag{1}$$

with three extra parameters describing the bilinear RPB coupling of the three generations of leptons to the u-type Higgs superfield. In addition there are three new parameters to describe the new soft supersymmetry breaking terms

$$V = V_{MSSM} + B_i \varepsilon_i \tilde{l}_i H_u. \tag{2}$$

All the six RPB parameters are determined within fairly tight limits from the observed neutrino masses and mixing angles, so that one has effectively no free parameters other than those of the MSSM [5,6]. Moreover the small neutrino masses ensure very small RPB parameters $\varepsilon_i$, so that all the predictions of superparticle production and decay down to the LSP remain essentially the same as in the MSSM. But the LSP is predicted to decay with the decay range and branching ratios determined in terms of the abovementioned RPB parameters. The details of these predictions may be found in ref [6]. We shall only mention here the two main features of LSP decay, which shall be used in our analysis. Firstly the main decay channels of the LSP ($\chi$) are $(A) \chi \to \tau^+ \tau^- \nu \ \& \ (B) \chi \to \bar{b} b \nu$, with branching ratios of about 0.3 and 0.6 respectively over the relatively light SUSY mass range of our interest. Secondly its decay range, $r_0 = c\tau$, is about 1 mm for $m_\chi \cong 100 GeV$ and goes down inversely as its mass thereafter.

The first feature implies a strong degradation of the canonical missing-$p_T$ signature of the R-parity conserving MSSM at the LHC. Nonetheless one can get viable leptonic signatures for superparticle production at LHC from their cascade decay via wino into a bino LSP in the mSUGRA model, where one has a 2:1 hierarchy between the wino and bino masses [7]. On the other hand in a more general MSSM the two masses can be of roughly similar size, in which case it will be hard to get a viable signature at the LHC.

We investigate here the signature of such a bilinear RPB SUSY model at an ultrahigh energy neutrino telescope like IceCube [8] or a water Cerenkov telescope of similar size in the Mediterranean [9]. To be definitive we shall consider three sets of MSSM spectra:

$I) m_\chi \cong 100 GeV, m_{\widetilde{W}} \cong m_{\widetilde{Z}} \cong m_{\widetilde{l}} \cong 120 GeV, m_{\widetilde{q}} \cong 150 GeV$

$II) m_\chi \cong 120 GeV, m_{\widetilde{W}} \cong m_{\widetilde{Z}} \cong m_{\widetilde{l}} \cong 140 GeV, m_{\widetilde{q}} \cong 200 GeV$

$III) m_\chi \cong 120 GeV, m_{\widetilde{W}} \cong m_{\widetilde{Z}} \cong m_{\widetilde{l}} \cong 250 GeV, m_{\widetilde{q}} \cong 300 GeV$.

Admittedly the sets I and II represent relatively light sparticle masses, chosen to give favourable signal cross-sections. However, for the same reason they represent the most favourable superparticle spectra from naturalness consideration. Moreover, these two cases will be hard to probe at LHC in the RPB SUSY model because of the degradation of the missing-$p_T$ as well as the $p_T$ of the leptons coming from the cascade decay. For the same reason the TeVatron limit on squark masses do not apply to them, while they satisfy all the LEP limits [10]. Thus they represent a very important region of the MSSM parameter space, to be probed at the UHE neutrino telescopes. On the other hand the set III represents relatively high wino, zino and slepton masses, like the typical mSUGRA spectrum for the electroweak sector. Thanks to the 2:1 mass hierarchy between the wino and the bino (LSP), this case can be probed at LHC via the leptonic signature[7].

The SUSY signals of our interest come from the CC and NC processes

$$vq \xrightarrow{\widetilde{W}} \widetilde{l}\widetilde{q}, \widetilde{l} \to l\chi \qquad (3)$$

$$vq \xrightarrow{\widetilde{Z}} \widetilde{v}\widetilde{q}, \widetilde{v} \to v\chi \qquad (4)$$

followed by the decay of $\chi$ into channels A or B above. The cross-sections can be easily obtained from the corresponding ones derived in [11,12] for electro-production. As in ref [11] we shall neglect mixings for chargino and neutralino states, which we expect to be good approximations as long as the higgsino states are reasonably heavy relative to the wino and zino. The different helicity contributions to the CC process (3) are given by

$$\frac{d\sigma_{CC}^{LL,RR}}{dt} = \frac{\pi\alpha^2}{Sin^4\theta_W} \frac{m_{\widetilde{W}}^2}{s(t - m_{\widetilde{W}}^2)^2} ; \frac{d\sigma_{CC}^{LR,RL}}{dt} = \frac{\pi\alpha^2}{Sin^4\theta_W} \frac{(tu - m_{\widetilde{l}}^2 m_{\widetilde{q}}^2)}{s^2(t - m_{\widetilde{W}}^2)^2} . \qquad (5)$$

The corresponding contributions to the NC process (4) are obtained by substituting the zino mass for the wino and multiplying by the coupling factor $G^q_{L,R}/Cos^4\theta_W$, where

$$G^q_L = (0.5 - e_q Sin^2\theta_W)^2 \ \& \ G^q_R = (e_q Sin^2\theta_W)^2. \tag{6}$$

The resulting signal cross-sections are obtained by convoluting these CC and NC cross-sections with the corresponding quark densities, i.e.

$$\sigma_{CC} = \int_{x_{min}}^{1} dx \int_{t_{max}}^{t_{min}} dt \left(\frac{\pi\alpha^2}{4Sin^4\theta_W}\right)\left[\frac{m^2_{\widetilde{W}}}{s(t-m^2_{\widetilde{W}})^2}\sum_i d_i(x) + \bar{d}_i(x) + \frac{(tu - m^2_{\tilde{l}} m^2_{\tilde{q}})}{s^2(t-m^2_{\widetilde{W}})^2}\sum_i u_i(x) + \bar{u}_i(x)\right] \tag{7}$$

$$\sigma_{NC} = \int_{x_{min}}^{1} dx \int_{t_{max}}^{t_{min}} dt \left(\frac{\pi\alpha^2}{2Sin^4\theta_W Cos^4\theta_W}\right) \times$$

$$\left[\frac{m^2_{\widetilde{Z}}}{s(t-m^2_{\widetilde{Z}})^2}\sum_i (G^u_L u_i(x) + G^u_R \bar{u}_i(x) + G^d_L d_i(x) + G^d_R \bar{d}_i(x)) + \frac{(tu - m^2_{\tilde{l}} m^2_{\tilde{q}})}{s^2(t-m^2_{\widetilde{Z}})^2}\sum_i (G^u_L \bar{u}_i(x) + G^u_R u_i(x) + G^d_L \bar{d}_i(x) + G^d_R d_i(x))\right] \tag{8}$$

where *i* is the generation index and

$$t_{min,max} = -(s - m^2_{\tilde{l}} - m^2_{\tilde{q}} \mp \sqrt{(s - m^2_{\tilde{l}} - m^2_{\tilde{q}})^2 - 4m^2_{\tilde{l}} m^2_{\tilde{q}}})/2; x_{min} = (m_{\tilde{l}} + m_{\tilde{q}})^2 / 2E_\nu m. \tag{9}$$

Due to the large squark and slepton masses the minimum x for the SUSY processes is at least four orders of magnitude larger than that for the SM CC processes like

$$\nu_\tau q \xrightarrow{W} \tau q', \tag{10}$$

and the strong rise of the sea quark densities at low x implies that the cross-sections for the SUSY processes (3,4) are suppressed by at least two orders of magnitude relative to this SM CC cross-section.

We have computed the SUSY signal cross-section from the CC and NC processes (3,4) using the CTEQ4L quark densities, setting the scale $Q^2$ = s. We have also checked that there is very little change in the result for alternative choices of scale or quark densities. Fig.1 shows the SUSY signal cross-sections for the MSSM spectra of set I, II and III over the UHE neutrino energy range $E_\nu$ = 10 -1000 PeV. Our cross-section for the set III matches with the corresponding cross-section computed recently [13], in the context of a different extension of the MSSM. For comparison we also show the leading order SM CC cross-section using the simple parametrisation of [14], i.e.

$$\sigma^{SM}_{CC} = 5.53 pb(E_\nu / GeV)^{0.363}. \tag{11}$$

The NLO correction along with uncertainty of quark densities can change this cross-section by about 30% [15]. The SM CC cross-section is indeed seen to be larger than the SUSY cross-sections of sets I-III by 2-3 orders of magnitude. It represents CC production of e, μ or τ. In particular the τ production process (10) is expected to have a spectacular double bang signature, resulting from the production and the hadronic decay vertices of τ, with a separation of ~ $10^2$ meters in the multi-PeV energy range of our interest [16]. This will constitute an important bench-mark for the collinear triple bang signature for SUSY events, discussed below, as they would both have similar detection efficiencies.

The multi-PeV LSP, produced by the CC and NC processes (3,4) is expected to decay mainly into the $\tau^+\tau\nu$ and $\bar{b}b\nu$ channels. The former decay leads to a collinear triple bang signature, coming form the production vertex of (3,4) and the hadronic decay vertices of the two taus, again with typical separations of ~ $10^2$ meters. For a quantitative analysis, we have performed a Monte Carlo simulation of LSP production via (3,4), followed by its decay, $\chi \to \tau^+\tau^-\nu$. Each event records the energies of the LSP as well as the two decay taus, and orders the latter according to their energies. Thus one gets the Lorentz boosted decay range ($r = r_0 E/m$) of the LSP along with those of the more and less energetic taus. They are shown in fig.2 for an incident neutrino energy of 20 PeV, since most of the signal events discussed below come from the lowest energy bin, $E_\nu = 20 \pm 10 PeV$. Note that the above LSP decay vertex is marked by the appearance of two collinear tau tracks, which may be hard to identify at the IceCube. Therefore we have added the LSP range to those of the two decay taus and shown the resulting effective decay ranges of the more and the less energetic taus relative to the production vertex . Thus they represent the ranges of the $2^{nd}$ and $3^{rd}$ collinear bangs relative to the $1^{st}$. We see that for the MSSM spectra I and II the LSP decay range is peaked at 100 meters, while the effective decay ranges of the two taus are peaked at 200 and 500 meters. The corresponding ranges for the set III are reduced by about half each. This is because the LSP carries most of the slepton energy for the spectra I and II as they have similar masses, while for III it carries about half the slepton energy. Note that all these ranges scale with the incident neutrino energy. Thus one can reduce the ranges by going to lower neutrino energy, which will in fact increase the signal rate.

It should be noted here that about 2/3$^{rd}$ of the signal cross-section comes from the CC process (3), which gives a multi-PeV charged lepton, collinear with the triple bang. This can be identified via one more bang for tau, a clear track for muon and showering for electron. This constitutes an additional distinctive feature of the above signal. This feature is even more important for the $\chi \to \bar{b}b\nu$ decay channel, which would by itself show up as a double bang event. The presence of the collinear multi-PeV charged lepton will give a third bang for tau, a clear track for muon or shower for electron. This can clearly distinguish this signal from the double bang events of the SM CC process (10), at least in the first two cases. In estimating the signal size below we shall add 2/3$^{rd}$ of the BR for this channel (40%) to the BR of the $\tau^+\tau^-\nu$ decay channel (30%), giving an effective BR of 70% for the SUSY signal.

It should be added here that the NC SUSY process (4) has also a distinct feature, in the sense that the first bang from the production vertex is followed by a clear gap of ≈ 100 meters, corresponding to the decay range of the LSP $\chi$. This can distinguish the signal from the double bang events of the SM CC process (10), which are connected by the tau track, provided one can identify this track. However this may be quite hard at the IceCube, as mentioned earlier. It should also be noted that the SUSY processes (3) and (4) have a second LSP, coming from the squark decay. The decay of this LSP ($\chi$) will give additional bang(s). Being a fragment of the nucleon target, however, the squark and its decay products carry less energies than those of the slepton. Our Monte Carlo simulation shows that the LSP ($\chi$) and the resulting $\tau$ leptons from the squark decay have energies, which are smaller than the energies of the corresponding particles from the slepton decay by about a factor of 5 each. Accordingly the Lorentz boosted decay ranges of these particles are smaller by about a factor of 5 each relative to the corresponding ones shown in Fig 2. So only the more energetic $\tau$ lepton from the squark decay has a range of about 100 meters from the production vertex, while the less energetic $\tau$ lepton and the LSP have decay ranges ≤ 50 meters from the production vertex. Hence only the more energetic $\tau$ lepton decay will give an extra bang, which can be resolved from that of the production vertex at the IceCube. One can easily check that including this extra bang from the squark decay $\tau$ does not lead to any significant enhancement of the signal BR. Therefore we shall keep the above mentioned effective BR of 70% for the SUSY signal in this simple analysis.

Finally we come to the bad news, i.e. the event rate. It is estimated by convoluting the above signal cross-section with the UHE neutrino flux, which is unfortunately clouded by a large uncertainty. The most popular choice is the so called Waxman-Bahcall (WB) flux, which assumes a common extragalactic source for the UHE neutrinos and cosmic ray protons [17]. It accelerates protons to UHE, which then interacts with the ambient photons to give pions, $p\gamma \to n\pi^+(p\pi^0), \pi^+ \to \nu_\mu \bar{\nu}_\mu \nu_e e^+$. The neutron escapes the confining magnetic field of the source along with the neutrinos from pion decay. They further assume the source to be optically thin, so that the neutron escapes without interaction with the ambient photons, and decays outside to give the CR proton. So the UHE neutrino and CR proton fluxes are correlated to one another. Finally they assume the typical $E^{-2}$ power law of a Fermi engine for the neutrino/CR spectra. The predicted neutrino flux, including all flavours of neutrino and antineutrino, is

$$J_\nu^{WB} = 6 \times 10^{-8} (E/GeV)^{-2} GeV^{-1} cm^{-2} s^{-1} sr^{-1}, \qquad (12)$$

which is equally distributed into the three flavours by neutrino oscillations. However the predicted normalisation has a rather large model dependent uncertainty. Besides the assumed power law of -2 has been questioned by many authors, who suggest to treat it instead as a free parameter [18]. Indeed as noted in [19], the AGASA and HiRes data both show a steepening of the UHE CR spectrum from $E^{-2}$ to $E^{-2.54}$ above $E = 3 \times 10^8 GeV$. This coincides with the change of CR composition from heavy nuclei to proton dominance, marking the dominance of the extragalactic component. Based on this observation these authors have used a WB type model to obtain a UHE neutrino flux from these CR data. They assume the above photo-meson production process to be dominated by the $\Delta^+$ resonance, so that the energy sharing between the pion and the neutron is kinematically determined. The pion is predicted to carry about 28% of the neutron energy and each of its decay neutrinos about 7% of the latter. Thus the CR proton flux above $E \approx 3 \times 10^8 GeV$ determines the neutrino flux above $E_\nu \approx 10^7 GeV$. Using the average of AGASA and HiRes CR fluxes for normalisation, they predict a total neutrino flux for all flavours [19]

$$J_\nu^{AH}(E) = 3.5 \times 10^{-3} (E/GeV)^{-2.54} GeV^{-1} cm^{-2} s^{-1} sr^{-1}. \qquad (13)$$

Apart from this there can be a contribution to the UHE neutrino flux from optically thick sources like the cores of active galactic nuclei (AGN), from which no nucleon can escape [21]. The only particles coming out of the abovementioned photo-pion production process in these sources are the neutrinos and photons resulting from the charged and neutral pion decays. Thus the UHE neutrino flux in this case is correlated to a UHE photon flux instead of CR. The photon flux cascades down to the GeV energy range in passage. Assuming the extragalactic component of the EGRET photon flux to be saturated by this contribution, gives an upper limit to this UHE neutrino flux, which is about thirty times larger than the WB flux at $E_\nu \approx 20 PeV$ [21].

Since the earth is opaque to the neutrinos in the $E_\nu \geq 10 PeV$ energy range of our interest [22], we shall only consider the down going neutrinos, covering a solid angle of $2\pi$ sr. Then the predicted number of signal events at IceCube is given by

$$N = 2\pi N_T T \int J_\nu(E_\nu) \sigma(E_\nu) dE_\nu ,\qquad(14)$$

where $N_T = 6 \times 10^{38}$ is the number of target nucleons in a km$^3$ size ice/water telescope and T is its total operation time, which will be taken as 15 years. Table 1 shows the expected number of signal events corresponding to the MSSM spectrum I along with those of the SM CC process (10) for the three UHE neutrino fluxes mentioned above. We have incorporated the abovementioned BR of 0.7 for the former and the flavour factor of 1/3 for the latter.

**Table 1**: Predicted number of events for the 15 years of IceCube operation. The number of signal events for SUSY spectra -II(III) are 1/2 (1/10) of those shown for SUSY-I.

| Neutrino Flux | Waxman-Bahcall | AGASA-HiRes | AGN Model |
|---|---|---|---|
| SM-τ (double bang) | 12 | 65 | ~ 360 |
| SUSY-I | 0.34 | 1.8 | ~ 10 |

Note that the number of signal events in the bilinear RPB SUSY model considered here is the same as the standard MSSM. But unlike the latter the RPB model has a distinctive signature at a UHE neutrino telescope, so that the viability of the signal is primarily determined by the number of signal events. However the expected number of signal events at IceCube, shown in Table 1, is admittedly too small to give a viable SUSY signal. Even the most optimistic prediction of ~ 10 events may at best be marginally viable after taking into account the detection efficiency of IceCube. Nonetheless it is encouraging to note that the IceCube will at least come within striking range of detecting this SUSY signal. We hope that the successful operation of IceCube along with a similar sized water Cerenkov telescope at the Mediterranean will lead to construction of a ~10 km$^3$ size UHE telescope in the future. Meanwhile the UHE neutrino flux would have been measured at the IceCube. So we hope that this telescope can effectively probe this SUSY signal at least for the MSSM spectra I and II. As mentioned before these two sets represent the most natural part of the SUSY parameter space, which may be hard to probe at LHC.

In summary, we have considered a predictive and well motivated extension of the MSSM, which has a distinctive signature at the UHE neutrino telescopes. Admittedly the expected event rate is at best marginal at the IceCube. But we hope that a ~10 km$^3$ UHE neutrino telescope in the future will be able to probe this signal more effectively, at least for a very important part of the SUSY parameter space.

We thank Manuel Drees, Raj Gandhi, Monoranjan Guchait, Francis Halzen, Sandip Pakvasa, Ricard Tomas and Thomas Weiler for discussions. This work was inspired by discussions at the Aspen Institute 2007 Neutrino Physics Programme. It was supported by Spanish grants FPA2005-01269, FPA2005-25348-E(MEC), SAB2005-0131 and European Commission Contract MRTN-CT-2004-503369. DPR was supported in part by BRNS(DAE) through the Raja Ramanna Fellowship and by MEC SAB2005-0131.

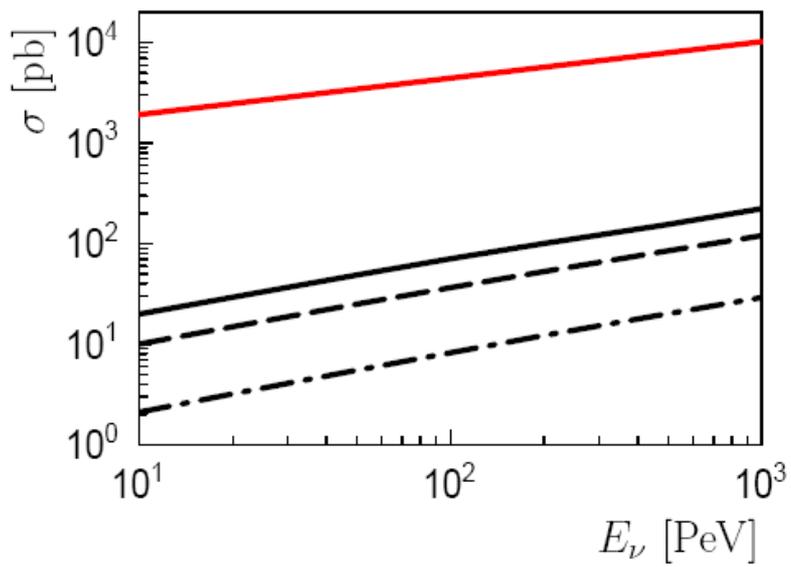

Fig1. Signal cross-sections for SUSY spectra I (solid), II (dashed) and III (dot-dashed) shown against the neutrino energy along with the SM CC cross-section (top line).

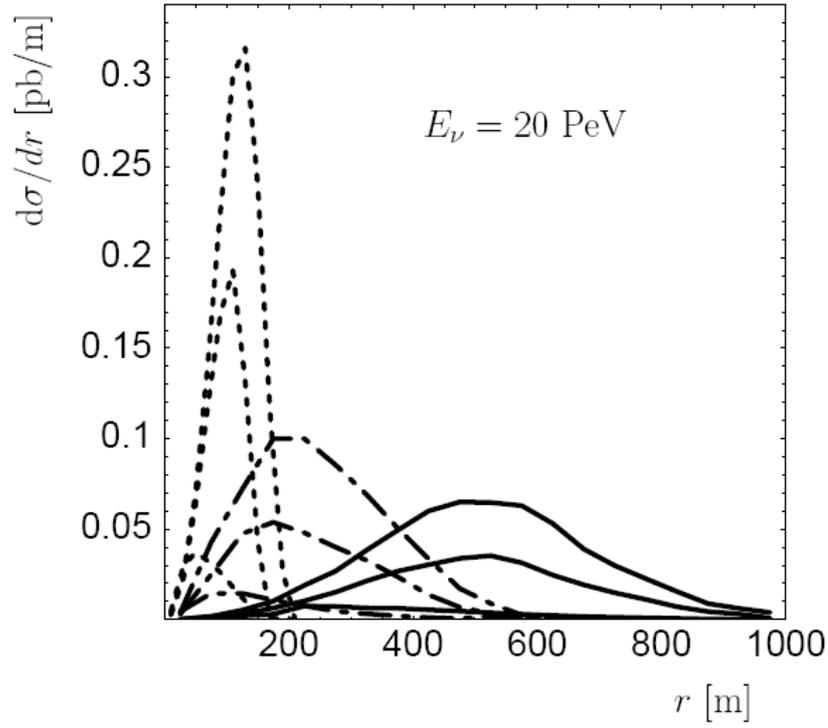

Fig2. The Lorentz boosted range of the LSP (dotted) shown along with those of the more (solid) and less (dot-dashed) energetic decay taus. The lines from top to bottom represent the SUSY spectra I, II and III respectively.